\documentclass[showpacs,preprintnumbers,amsmath,amssymb,prc,12pt]{revtex4}

\usepackage{graphicx}
\usepackage{amssymb}
\usepackage{epstopdf}
\preprint{NSF-KITP-06-109}

\begin{document}

\def\aprge{\buildrel > \over {_{\sim}}}
\def\aprle{\buildrel < \over {_{\sim}}}

\def\etal{{\it et.~al.}}
\def\ie{{\it i.e.}}
\def\eg{{\it e.g.}}

\def\bwt{\begin{widetext}}
\def\ewt{\end{widetext}}
\def\be{\begin{equation}}
\def\ee{\end{equation}}
\def\bea{\begin{eqnarray}}
\def\eea{\end{eqnarray}}
\def\bean{\begin{eqnarray*}}
\def\eean{\end{eqnarray*}}
\def\bary{\begin{array}}
\def\eary{\end{array}}
\def\bi{\bibitem}
\def\bit{\begin{itemize}}
\def\eit{\end{itemize}}

\def\lan{\langle}
\def\ran{\rangle}
\def\lra{\leftrightarrow}
\def\la{\leftarrow}
\def\ra{\rightarrow}
\def\dash{\mbox{-}}
\def\ol{\overline}

\def\ub{\ol{u}}
\def\db{\ol{d}}
\def\sb{\ol{s}}
\def\cb{\ol{c}}

\def\re{\rm Re}
\def\im{\rm Im}

\def \b{{\cal B}}
\def \ca{{\cal A}}
\def \ko{K^0}
\def \ok{\overline{K}^0}
\def \s{\sqrt{2}}
\def \st{\sqrt{3}}
\def \sx{\sqrt{6}}

\title{\Large{\bf On the Radiatively Induced Lorentz and \\ CPT Violating  Chern-Simons Term }}
\author{Yong-Liang Ma\footnote{Current address: Institute for Theoretical Physics, University of Tuebingen,
   D-72076 Tuebingen, Germany}  and Yue-Liang Wu}
\address{Kavli Institute for Theoretical Physics China (KITPC) \\ Institute of Theoretical Physics, Chinese Academy of Sciences, Beijing
100080, China \\
Kavli Institute for Theoretical Physics \\  University of California, Santa Barbara, CA 93106}
\date{\today}
\begin{abstract}
The radiatively induced Lorentz and CPT violating Chern-Simons terms
in QED is calculated based on the recently developed loop
regularization method [Y.L. Wu, Int.J.Mod.Phys.{\bf A18} (2003)
5363, hep-th/0209021; Y.L. Wu, Mod.Phys.Lett.{\bf A19} (2004) 2191,
hep-th/0311082] for quantum field theories. It enables us to make
general comments on the various results in literature and obtain a
consistent result when simultaneously combining the evaluation for
the chiral anomaly which has a unique form once the vector current
is kept conserved.
\end{abstract}
 \maketitle

Quantum field theory based on the Lorentz and CPT invariance has got
great triumphes. It is believed that CPT invariance is a fundamental
property of a relativistic point particle. While it is also thought
that, quantum field theory may not be an underlying theory but an
effective theory describing physics phenomena below some
characteristic energy scales\cite{weinberg-book}. One of the
interesting extensions in quantum field theory has been payed on the
investigation of the Lorentz and CPT violation\cite{Carroll90,LCPT3,
Coleman99}. For this case, in order to interpret the success of
quantum field theory, the coupling constant of the extended terms is
understood to be suppressed by a scale such as
$m_W/M_P\simeq10^{-17}$. The Lorentz violation was shown to be
consistent with string theory in which the Lorentz invariance is
violated spontanously\cite{string}. An intrinsically Lorentz
non-invariant\cite{NC-1} can also appear in the noncommutative field
theories\cite{NC2,NC-2} as a deduction of string theory.

Naively, one can simply add Lorentz non-invariant terms to get a
Lorentz violating quantum field theory. In general, the Lorentz
symmetry can be broken down explicitly or spontaneously. It can also
be induced from other terms of a large theory through quantum
corrections. In this note, we will focus on the latter case and
devote to investigate the Chern-Simons term
${1\over2}k_\mu\epsilon^{\mu\nu\alpha\beta}F_{\nu\alpha}A_\beta$
(with $k_\mu$ as a constant vector) from the extended
QED(EQED)\cite{Jakiw99}
\begin{eqnarray}
{\cal
L}_f&=&\bar{\psi}(iD\hspace{-0.27cm}\slash-m-b\hspace{-0.2cm}\slash\gamma_5)\psi\label{EQED}
\end{eqnarray}
with $D_\mu=\partial_\mu-iA_\mu$. Here $m$ is the fermion mass and
$b_\mu$ is a constant vector. Our goal is to determine the relation
between $k_\mu$ and $b_\mu$ and discuss the ambiguities in various
calculations.

The relation between $k_\mu$ and $b_\mu$ has been discussed by many
groups, an infinity class may be resulted relying on various regularization schemes.
Here we mainly comment on three interesting results:
\begin{enumerate}
  \item $k_\mu=0$ which was resulted based on the gauge invariance of the axial-vector
  current $j_{\mu 5}$ \cite{Coleman99}. The same conclusion was also
  reached in \cite{LCPT3} by using Pauli-Villars regularization in the case of
  $m\rightarrow\infty$, and in \cite{MPV} by considering the gauge
invariance and the conservation of vector Ward identity of the
triangle diagram in the massive case $m\neq 0$, as well as in
\cite{GB} via a consistent analysis based on dimensional
regularization.
  \item $k_\mu=\frac{3}{16\pi^2}b_\mu$ which was obtained via evaluating the relative amplitudes in a nonperturbative
  formulation\cite{Jakiw99} for the case $m\neq 0$. The same result was
  yielded in ref. \cite{Chung99} via the derivative expansion with the use of dimensional regularization and also reached
  in ref. \cite{Victoria} by keeping the full $b_\mu$ dependence in case of $m^2\geq -b^2$.
  \item $k_\mu=-\frac{1}{16\pi^2}b_\mu$ which was concluded in \cite{Victoria} for the case of
  $m=0$.
\end{enumerate}

In this note, we shall comment on the above results from an alternative calculation based on the
recently developed Loop regularization\cite{LR} which has
successfully been applied to obtain a consistent result for the chiral anomaly\cite{Ma05}.
Consequently, we arrive at the following relations
\begin{eqnarray}
k_\mu=0 \label{k-m-infinity}
\end{eqnarray}
for the case $m \neq 0$ , which reproduces the result obtained in
\cite{LCPT3,Coleman99,MPV,GB}, and
\begin{eqnarray}
k_\mu=\frac{1}{4\pi^2}b_\mu \label{k-m-zero}
\end{eqnarray}
for the case $m = 0$.

We will also show that the results $k_\mu = \frac{3}{16\pi^2} b_\mu$
for the case $m=0$ and $k_\mu = -\frac{1}{16\pi^2} b_\mu$ for the
case $m\neq 0$ are resulted when making an alternative treatment for
the linearly divergent integrals. Nevertheless, such an alternative
treatment cannot obtain the standard form for the chiral triangle
anomaly\cite{A1,A2} which has been found to be unique when the conservation of the
vector current is imposed and been tested by the experiment in the
process $\pi^0 \to \gamma \gamma$.

We begin our considerations following the general description in ref.\cite{Jakiw99}
and calculate the amplitude of three point Green functions with zero
momentum for the axial-vector current. For this purpose, it
only needs to keep the leading-order of $b_{\mu}$, namely
\begin{eqnarray}
\Pi^{\mu\nu}_b(q)&=&b_\lambda[\Pi^{\mu\nu\lambda,(1)}(q)+\Pi^{\mu\nu\lambda,(2)}(q)]\label{Pi-b}
\end{eqnarray}
where
\begin{eqnarray}
\Pi^{\mu\nu\lambda,(1)}(q) & = & \int\frac{d^4p}{(2\pi)^4}{\rm
Tr}\{\gamma_\mu \frac{1}{p\hspace{-0.2cm}\slash-m}\gamma_\nu
\frac{1}{p\hspace{-0.2cm}\slash+q\hspace{-0.2cm}\slash-m}\gamma_\lambda\gamma_5
\frac{1}{p\hspace{-0.2cm}\slash+q\hspace{-0.2cm}\slash-m}\}\\
\Pi^{\mu\nu\lambda,(2)}(q) & = & \int\frac{d^4p}{(2\pi)^4}{\rm
Tr}\{\gamma_\mu
\frac{1}{p\hspace{-0.2cm}\slash-m}\gamma_\lambda\gamma_5\frac{1}{p\hspace{-0.2cm}\slash-m}\gamma_\nu
\frac{1}{p\hspace{-0.2cm}\slash+q\hspace{-0.2cm}\slash-m}\}
\end{eqnarray}

To evaluate two amplitudes $\Pi^{\mu\nu\lambda,(1)}$ and
$\Pi^{\mu\nu\lambda,(2)}$, a regularization scheme should be adopted
as they are both divergent integrals. Also as they involve
$\gamma_5$ in the amplitudes, the dimensional regularization is not
suitable. Here we shall use the loop regularization(LR)\cite{LR}
which is realized in the original dimension without changing the field content of the original
theory. This is different from the Pauli-Villars(PV) regularization
which introduces some superheavy particles as regulator fields. As a
consequence, the LR scheme, unlike the PV scheme, can be applied to
the Non-abelian gauge theory without spoiling any symmetries
including the non-abelian gauge symmetry though two mass scales are
intrinsically introduced. Those two mass scales play the role of
characterizing energy scale ($M_c$) and sliding energy scale
($\mu_s$). It has been shown that the LR scheme can be applied
not only to the underlying gauge theories\cite{LR}, but also to the effective
quantum field theory for understanding the dynamically spontaneous symmetry
breaking\cite{DW} and the chiral theory for clarifying the possible
ambiguities appearing in chiral anomaly\cite{Ma05}.

The prescription of loop regularization is simple\cite{LR}: Firstly
evaluating the Feynman integrals into irreducible loop
integrals(ILIs). The ILIs  at one-loop level have the following
general forms
\begin{eqnarray}
I_{-2a}&=&\int\frac{d^4k}{(2\pi)^4}\frac{1}{(k^2-M^2)^{2+a}}\\
I_{-2a~\mu\nu}&=&\int\frac{d^4k}{(2\pi)^4}\frac{k_\mu
k_\nu}{(k^2-M^2)^{3+a}},\ a=-1,0,1,2,\cdots \nonumber \label{ILIs}
\end{eqnarray}
Then replacing the integration variable $k^2$ and integration
measure $\int\frac{d^4k}{(2\pi)^4}$ by the regularized ones
\begin{eqnarray}
& & k^2\rightarrow[k^2]_l\equiv k^2-M_l^2, \nonumber \\
& & \int\frac{d^4k}{(2\pi)^4}\rightarrow
\int[\frac{d^4k}{(2\pi)^4}]_l\equiv\lim_{N,M_i^2}
\sum_{l=0}^Nc_l^N\int\frac{d^4k}{(2\pi)^4}\label{precedure}
\end{eqnarray}
where $c_l^N$ are the coefficients determined by the following
conditions.
\begin{eqnarray}
\lim_{N,M_i^2}\sum_{l=0}^Nc_l^N(M_l^2)^n=0, \quad c_0^N=1
\label{condition}
\end{eqnarray}
with $ i=0,1,\cdots,N~~\mbox{and} ~~n=0,1,\cdots $. Note that taking
$M_i\to \infty$ and $\mu_s=0$, the initial integral is recovered.
Also taking $M_i$ and $N$ to be infinity, so that the regularized
theory becomes regulator independent. It has been shown that such
regularized ILIs satisfy a set of consistent conditions\cite{LR}
\begin{eqnarray}
& & I^R_{2\mu\nu}= {1\over2}g_{\mu\nu}I^R_2,\quad I^R_{0\mu\nu}= {1\over4}g_{\mu\nu}I^R_0, \quad \cdots
\end{eqnarray}
which ensure the gauge and Lorentz invariance. For the simple form
of  the regulator masses $M_l = \mu_s + M_R\ l$ ($l= 0,1, \cdots$),
the coefficients $c_l^N$ is found to be $ c_l^N = (-1)^l
\frac{N!}{(N-l)!\ l!}$.  The integrals $I_2^R$ and $I_0^R$ are then given by the following explicit forms \cite{LR}
\begin{eqnarray}
I_2^R&=&-\frac{i}{16\pi^2}\{M_c^2-\mu^2[\ln\frac{M_c^2}{\mu^2}-\gamma_\omega+1+
y_2(\frac{\mu^2}{M_c^2})]\}\label{I2R} \nonumber \\
I_0^R&=&\frac{i}{16\pi^2}[\ln\frac{M_c^2}{\mu^2}-\gamma_\omega
+y_0(\frac{\mu^2}{M_c^2})]\label{I0R}
\end{eqnarray}
with $ \mu^2 = \mu_s^2 + M^2$, $\gamma_w = \gamma_E = 0.5772\cdots$,
and
\begin{eqnarray}
& & y_0 (x) = \int_0^x d \sigma\ \frac{1 - e^{-\sigma} }{\sigma},
\quad  y_1 (x)  = \frac{e^{-x} - 1 + x}{x} \nonumber \\
& & y_2(x) = y_0(x) - y_1(x), \quad    M_c^2 = \lim_{N,M_R} M_R^2/\ln N \label{y-Function}
\end{eqnarray}
Here $\mu_s $ sets an IR `cutoff' at $M^2 =0$ and $M_c$ provides an
UV `cutoff'. For renormalizable quantum field theories, $M_c$ can be
taken to be infinity $(M_c\rightarrow\infty)$. $\mu_s$ can safely
runs to $\mu_s=0$ in a theory without infrared divergence.

Let us now calculate the amplitude in eq.(\ref{Pi-b}). We shall
first consider $\Pi^{\mu\nu\lambda,(1)}$. After using the following
convention and relation
\begin{eqnarray}
&&{\rm
Tr}\{\gamma_5\gamma_\mu\gamma_\nu\gamma_\alpha\gamma_\beta\}=-4i\epsilon_{\mu\nu\alpha\beta},~~~~\epsilon_{0123}=1\\
&&2(p+q)\cdot p=(p+q)^2+p^2-q^2
\end{eqnarray}
The amplitude $\Pi^{\mu\nu\lambda,(1)}$ can be expressed in terms of
the divergent part $\Pi^{\mu\nu\lambda,(1)}_D$ and the convergent
part $\Pi^{\mu\nu\lambda,(1)}_C$
\begin{eqnarray}
\Pi^{\mu\nu\lambda,(1)} = \Pi^{\mu\nu\lambda,(1)}_D+\Pi^{\mu\nu\lambda,(1)}_{C}
\end{eqnarray}
For the convergent part $\Pi^{\mu\nu\lambda,(1)}_{C}$, its
regularized amplitude is found in the limit $M_c\to \infty$ to be
\begin{eqnarray}
\Pi^{\lambda\mu\nu,(1)R}_C & = &
\frac{i}{4\pi^2}\epsilon_{\lambda\alpha\nu\mu}q_\alpha \int_0^1dx
x(1-x)\frac{q^2}{(m^2+\mu_s^2)-x(1-x)q^2}
\end{eqnarray}
with $x$ the Feynman parameter. For the divergent part, it can be decomposed into the
logarithmically divergent part $\Pi^{\mu\nu\lambda,(1)}_{0}$ and
linearly divergent part $\Pi^{\mu\nu\lambda,(1)}_{1}$
\begin{eqnarray}
\Pi^{\lambda\mu\nu,(1)}_D & = & \Pi^{\mu\nu\lambda,(1)}_{0}+\Pi^{\mu\nu\lambda,(1)}_{1}\\
\Pi^{\mu\nu\lambda,(1)}_{0} & = &  4\int\frac{d^4p}{(2\pi)^4}\frac{-1}{[(p+q)^2-m^2]^2(p^2-m^2)}\nonumber\\
&&\times\bigg\{\epsilon_{\lambda\alpha\nu\beta}(p+q)_\alpha p_\beta
(p+q)_\mu + \epsilon_{\lambda\alpha\beta\mu}(p+q)_{\alpha}p_\beta(p+q)_\nu \nonumber\\
&&+\epsilon_{\lambda\nu\beta\rho}(p+q)_\mu p_\beta(p+q)_\rho
+\epsilon_{\lambda\beta\mu\rho}(p+q)_\nu p_\beta(p+q)_\rho\nonumber\\
&&+\epsilon_{\alpha\nu\beta\mu}(p+q)_\alpha k_\beta(p+q)_\lambda
-\epsilon_{\nu\mu\beta\rho}(p+q)_\lambda p_\beta(p+q)_\rho\bigg\}\nonumber\\
& & - 4\epsilon_{\mu\nu\lambda\alpha}\int\frac{d^4p}{(2\pi)^4}\{\frac{q_\alpha}{[(p+q)^2-m^2][p^2-m^2]}\}\label{log-div}\\
\Pi^{\mu\nu\lambda,(1)}_{1} & = &
4\epsilon_{\lambda\alpha\nu\mu}\int\frac{d^4p}{(2\pi)^4}\frac{(p+q)_\alpha}{[(p+q)^2-m^2]^2}\label{LinearTerm}
\end{eqnarray}

After applying the LR scheme to the divergent integrals, the
regularized divergent part $\Pi^{\lambda\mu\nu,(1)R}_D$ becomes well
defined and can be evaluated consistently. Explicit
calculation gives
\begin{eqnarray}
\Pi^{\mu\nu\lambda,(1)R}_{0} & = & 0
\end{eqnarray}
due to cancellation among the logarithmically divergent terms. For the reguralized linearly divergent part
$\Pi^{\mu\nu\lambda,(1)R}_{1}$, we have
\begin{eqnarray}
\Pi^{\mu\nu\lambda,(1)R }_{1} & = &
\lim_{N,M_l^2}\sum_{l=0}^Nc_l^N\int\frac{d^4p}{(2\pi)^4}\frac{p_\mu
+ q_{\mu}}{[(p+q)^2-\hat{M}_l^2]^2} \nonumber
 \\
& = & \lim_{N,M_l^2}\sum_{l=0}^Nc_l^N\int\frac{d^4p}{(2\pi)^4}\frac{p_\mu}{[p^2-\hat{M}_l^2]^2}\nonumber\\
& &-\lim_{N,M_l^2}\sum_{l=0}^Nc_l^N\frac{i}{32\pi^2}q_\mu
 \label{surface-LR}
\end{eqnarray}
with $\hat{M}_l^2=M_l^2+m^2$. Here we have used the following
relation for the linearly divergent integral\cite{surfaceterm}
\begin{eqnarray}
\int\frac{d^4p}{(2\pi)^4}\frac{p_\mu}{[(p+q)^2-m^2]^2} & = &
\int\frac{d^4p}{(2\pi)^4}\frac{(p-q)_\mu}{[p^2-m^2]^2}
-\frac{i}{32\pi^2}q_\mu\label{surface}
\end{eqnarray}
By using the consistent condition in eq. (\ref{condition}) with
$n=0$, the second term of eq. (\ref{surface-LR}) vanishes. Its first
term also manifestly vanishes due to odd property of its integrant
function. Thus the regularized linearly divergent integral also vanishes
\begin{eqnarray}
\Pi^{\mu\nu\lambda,(1)R}_{1} & = & 0
\end{eqnarray}
As a consequence, the regularized amplitude
$\Pi^{\mu\nu\lambda,(1)R}$ is given in the limit $M_c\to \infty$ by
\begin{eqnarray}
\Pi^{\mu\nu\lambda,(1)R} & = & \Pi^{\mu\nu\lambda,(1)R}_C+\Pi^{\mu\nu\lambda,(1)R}_0+\Pi^{\mu\nu\lambda,(1)R}_1 \\
& = &
\frac{i}{4\pi^2}\epsilon_{\mu\nu\lambda\alpha}q_\alpha\int_0^1dx
\frac{-x(1-x)q^2}{(m^2+\mu_s^2)-x(1-x)q^2}\nonumber
\end{eqnarray}

The calculation for the second part $\Pi^{\mu\nu\lambda,(2)}$ is
almost the same as $\Pi^{\mu\nu\lambda,(1)}$. The only difference is
that there is no linear divergent integral in
$\Pi^{\mu\nu\lambda,(2)}$. After making a proper rearrangement for
the various terms, we have
\begin{eqnarray}
\Pi^{\mu\nu\lambda,(2)R} & = & \Pi^{\mu\nu\lambda,(1)R}
\end{eqnarray}
The full regularized amplitude of $\Pi^{\mu\nu\lambda}$ is given by
\begin{eqnarray}
\Pi^{\mu\nu\lambda, R} & = & \Pi^{\mu\nu\lambda,(1)R} +
\Pi^{\mu\nu\lambda,(2)R} \label{T-final3} \\
& = &
\frac{i}{2\pi^2}\epsilon_{\mu\nu\lambda\alpha}q_\alpha\int_0^1dx
\frac{-x(1-x)q^2}{(m^2+\mu_s^2)-x(1-x)q^2}\nonumber
\end{eqnarray}
As EQED is known to be free of IR divergence, one can
safely set $\mu_s = 0$. We then arrive at  the results
\begin{eqnarray*}
k_\mu & = & \left\{
        \begin{array}{ll}
          0, & \hbox{$m\neq 0$;} \\
          \frac{1}{4\pi^2}b_\mu, & \hbox{$m=0$.}
        \end{array}
      \right.
\end{eqnarray*}
which has been emphasized  in  eqs.(\ref{k-m-infinity}) and
(\ref{k-m-zero}). Thus the solution based on LR scheme\cite{LR}
confirms the statement in refs.\cite{LCPT3,Coleman99,MPV,GB} for the
case $m\neq 0$. It also becomes clear that for the massless case or
only when the mass is much less than the momentum of currents, a
non-zero Chern-Simons term which violates Lorentz invariance can be
induced. It was also shown in \cite{MPV} that the result in the
massless case is equal to the one with a finite mass minus
$1/(4\pi^2) b_\mu$, a sign difference is caused by the definition.
The result based on the Schwinger proper-time method was found to be
smaller by a factor of two in \cite{CCG}, its possible ambiguity may
arise from the treatment on the limit $\lim_{x\to 0} x_\mu x_\nu/x^2
= C g_{\mu\nu}$ with $C$ arbitrary in general and the value $C=1/4$
was taken in \cite{CCG} for four dimensions.

In fact, it has been noticed in ref.\cite{Ma05} that for the chiral
anomaly of triangle diagrams, when the fermion mass is much larger
than the momentum square of the currents, it becomes anomaly free.
Only when the fermion mass is much less than the momentum square of
the currents, or in the massless case, the standard form of anomaly
can be yielded.

We now comment on the other two results based on the LR scheme with
paying attention to the treatment for the linearly divergent
integrals. In general, one may first use the relation
eq.(\ref{surface}) and then apply the LR prescription only to the
logarithmically divergent part. If doing so, the finite surface term
would be survived and the full renormalized amplitude becomes
\begin{eqnarray}
\Pi^{\mu\nu\lambda, R} & = &
\frac{i}{2\pi^2}\epsilon_{\mu\nu\lambda\alpha}q_\alpha\int_0^1dx
\frac{-x(1-x)q^2}{(m^2+\mu_s^2)-x(1-x)q^2}\nonumber \\
& & - \frac{i}{8\pi^2}\epsilon_{\mu\nu\lambda\alpha}q_\alpha
 \label{T-final2}
\end{eqnarray}
Taking $\mu_s = 0$, we obtain the following two relations
\begin{eqnarray}
k_\mu & = & \left\{
        \begin{array}{ll}
         \frac{3}{16\pi^2}b_\mu, & \hbox{$m=0$;} \\
          - \frac{1}{16\pi^2}b_\mu, & \hbox{$m\neq 0$.}
        \end{array}
      \right.
\end{eqnarray}
which coincide with two results yielded in ref. \cite{Jakiw99} and
ref.\cite{Victoria} but with opposite mass limits.

As mentioned in the introduction, the Chern-Simons term was also
evaluated in\cite{Chung99} by adopting the derivative expansion with
the use of dimensional regularization and the same result as
ref.\cite{Jakiw99} was arrived. Nevertheless, in that derivation,
the gamma matrix and momentum integral were treated separately.
Namely, the trace for gamma matrices was treated in four dimensions
while the momentum integral is in $d$ dimensions. Such a treatment
is not exact as some of the gamma matrices may live in $d$
dimensions when they contract with the integration momentum. In
fact, it is already known that $\gamma_5$ is an intrinsically four
dimensional object and its definition in dimensional regularization
is ambiguous. Such an ambiguity may cause some ambiguities in the
direct calculations of chiral anomaly\cite{Ma05}. For the
calculation in ref. \cite{Victoria}, a symmetric momentum integral
is adopted. However, this treatment is not exact in some cases as
mentioned in ref. \cite{Ma05}.

The question then becomes which procedure is the right one for applying
the prescription of LR scheme. To answer this
question, it is helpful to check the well-known chiral anomaly of
triangle diagrams. It is not difficult to find that for the linearly
divergent integrals if adopting the relation eq.(\ref{surface}) to
reduce a linear divergence to a logarithmical divergence, and
applying the LR prescription only to the logarithmical divergent
integrals, then the resulting vector currents become not conserved
due to anomaly term, meanwhile the axial-vector current also gets a
corresponding additional anomaly term. As a standard procedure, when one makes
a redefinition to keep the vector current conserved, the
corresponding additional anomaly term in the axial-vector current
simultaneously  disappears. As a consequence, the final result is equivalent to drop away the
finite term in the relation eq.(\ref{surface}) for a linearly
divergent integral. That means we shall adopt the initial prescription of LR
scheme, namely apply the LR scheme directly to the linearly divergent integrals. This
should be the case as the LR scheme preserves translational symmetry.

From the above analyzes, we can draw the conclusion that the results
$k_{\mu} = 0$ for $m\neq 0$ and $k_{\mu} = \frac{1}{4\pi^2}b_\mu$
for $m=0$ are consistent and unique solutions based on the loop
regularization method. Note that when directly evaluating the
diagrams with zero momentum for the axial-vector current, the
additional anomaly term in eq.(27) does not violate vector current
conservation law. It is because of this reason that causes
difficulty to judge which procedure is more reliable without
combining the calculations of chiral anomaly which has a unique
form.

In conclusion, based on the recently developed loop regularization
method\cite{LR}, the ambiguities in calculating the radiatively
induced Lorentz and CPT violating Chern-Simons term in EQED can be
clarified when simultaneously combining the calculations of chiral
anomaly which has well been understood from the loop regularization
calculation\cite{Ma05}.

\acknowledgments

\label{ACK}

The authors would like to thank R. Jackiw for his suggestion of performing the calculation based on
loop regularization. This work was supported in part by the key projects of
Chinese Academy of Sciences, the National Science Foundation of
China (NSFC) under the grant 10475105, 10491306. This research was also supported in part by the National Science Foundation under Grant No. PHY99-07949.


\end{document}